\documentclass[nofootinbib,notitlepage,superscriptaddress,10pt,aps,pra,twocolumn]{revtex4-1}
\usepackage[utf8x]{inputenc}
\usepackage{amsmath}
\usepackage{amssymb}
\usepackage{graphicx}
\usepackage[normalem]{ulem}
\usepackage{epsf,epsfig}
\usepackage{psfrag}
\usepackage{xcolor}
\usepackage{hyperref}

\begin{document}

\title{Linking loop quantum gravity quantization ambiguities with phenomenology}

\author{Suddhasattwa Brahma}
\affiliation{Center for Field Theory and Particle Physics, Fudan University, 200433 Shanghai, China}

\author{Michele Ronco}
\affiliation{Dipartimento di Fisica, Universit\`a di Roma ``La Sapienza", P.le A. Moro 2, 00185 Roma, Italy}
\affiliation{INFN, Sez.~Roma1, P.le A. Moro 2, 00185 Roma, Italy}

\author{Giovanni Amelino-Camelia}
\affiliation{Dipartimento di Fisica, Universit\`a di Roma ``La Sapienza", P.le A. Moro 2, 00185 Roma, Italy}
\affiliation{INFN, Sez.~Roma1, P.le A. Moro 2, 00185 Roma, Italy}

\author{Antonino Marcian\`o}
\affiliation{Center for Field Theory and Particle Physics, Fudan University, 200433 Shanghai, China}

\begin{abstract}
Fundamental quantum gravity theories are known to be notoriously difficult to extract viable testable predictions out of. In this paper, we aim to incorporate putative quantum corrections coming from loop quantum gravity in deriving modified dispersion relations for particles on a deformed Minkowski spacetime. We show how different choices of the Immirzi parameter can, in some cases, serendipitously lead to different outcomes for such modifications, depending on the quantization scheme chosen. This allows one to differentiate between these quantization choices via testable phenomenological predictions.
\end{abstract}

\maketitle

\section{Introduction}
The most interesting outcome from several models in loop quantum gravity (LQG) recently has been that of resolution of classical singularities \cite{BojowaldLQC,GPSchwarzschild,CGHSVacuum,LQGGowdyLRS}. In spherically symmetric LQG, the Schwarzschild black hole singularity is replaced by an `effectively' Euclidean region inside the black hole, commonly referred to as `signature-change' in LQG \cite{CovSpherSymm, WDWSS}. (A similar result is available for the $(1+1)-$dimensional CGHS black hole solution \cite{CovCGHS}.) Similarly, the space-like cosmological big bang singularity gets resolved on loop quantization (see \cite{MathStrLQC, BojowaldLQCreview} for a recent review of the topic). The crucial ingredient used for this mechanism is to regularize the curvature operator in LQG in a specified manner, whereby it is written in terms of holonomies (or parallel transports of the connections). (For a different mechanism in LQG employed for singularity-resolution using inverse-triad corrections, see \cite{LQGNoSingInvTr}.) Effectively, this procedure, called `polymerization', implies that one replaces connections by rigorously derived functions of it\footnote{Typically, they are replaced by bounded functions which, indeed, plays a crucial role in singularity resolution. However, for naive choices of representations of a non-compact group, for the self dual variables, this is not always true as we shall see in this article.}. We shall refer to incorporating such functions as `holonomy corrections' in this article.

All these results are pointing towards an emergence of non-(pseudo)Riemannian structures in models of LQG. Such quantum spacetime structures challenge our accepted notions of covariance and give rise to deformations in the algebra of hypersurface deformations\footnote{Here, the word `deformation' is used to mean two different things, which should be distinguishable from the context in which it is used.} \cite{DeformedGR2}. Modified constraint algebras have been recently explored also in multiscale theories \citep{defhdamult}, but no signature change was found. This provides further evidence that, as we will discuss here, signature change is a characteristic feature of LQG, intimately related to singularity resolution as mentioned above. In fact, in the particular case of LQG, holonomy corrections lead to a specific modification of the structure functions arising in the Dirac constraint algebra, whereby leading to signature change \cite{CovSpherSymm, CovGen}, as long as one works with the real-valued Ashtekar-Barbero connection as is more prevalent in the community. What happens when using the self dual Ashtekar variables is less clear, although some recent evidence points towards an \textit{undeformed} algebra \cite{SDSS}. However, this depends on the way in which one chooses to implement the holonomy corrections in the self dual case. We clear up this point about the algebra of the quantum-corrected constraints, while using the self dual variables, in some detail in this paper.

Covariance, coupled with the assumption of a classical underlying spacetime, implies that we can have higher derivative terms in an effective theory of gravity (obviously, in addition to the Einstein-Hilbert term). Such perturbative quantum corrections from higher curvature actions are suppressed by extra factors of the Planck mass $M_{Pl}$ whereas non-perturbative ones arising from LQG can easily avoid such restrictions. Thus, to get effects which are not as small as these, we need to turn to more specific corrections coming from a particular theory of quantum gravity such as LQG. However, covariance is a strong consistency condition by itself, and cannot thus be arbitrarily deformed (or worse, \textit{violated}) in a quantum theory. For details for conditions on a generally covariant quantization of background independent theories, refer to \cite{CovSpherSymm, CovGowdy}. Consistent deformations of the hypersurface deformation algebra, arising from LQG,  have exciting phenomenological consequences. It is the natural generalization of the local Poincar\'e algebra, where one has arbitrary coordinate transformations (or the infinite dimensional diffeomorphism symmetry) as the fundamental symmetry group of general relativity. However, since the Poincar\'e algebra describes symmetries of Minkowski spacetime, it can be derived as a special case from the full hypersurface deformation algebra in a systematic manner. It has also been shown recently that the deformations arising in the hypersurface deformation algebra can be related to deformations in the local Poincar\'e algebra, in the particular case of spherical symmetry \cite{DeformedGR1}. Although a direct relationship between LQG and noncommutative spacetimes has been known for some time in lower dimensions\citep{LQGkP3D}, \cite{LQGkPoincare} is the first work towards relating LQG deformations to a noncommutative $\kappa$-Minkowski background enjoying $\kappa-$Poincar\'e deformed symmetries, in the flat limit.

In this work, we focus on another phenomenological aspect of such deformations arising from LQG, namely the relationship between the energy and momentum of elementary particles on (deformed) Minkowski spacetimes. The familiar dispersion relation $E= m^2 + p^2$ is then modified by the presence of such deformations and form a basis of testable predictions for effects arising from LQG. As already mentioned, most approaches to deriving the modified dispersion relation (MDR) come from considering Planck-suppressed Lorentz violating terms in an effective theory. An exception to this is doubly special relativity (DSR) where one postulates two different invariant scales instead of just the speed of light \citep{DSR1,DSR2,DSR3}. Our work provides an example of derivation of a MDR from a fundamental quantum gravity theory, which is similar in spirit to DSR. However, our results also aim to use this phenomenological prediction as a tool to discriminate between different approaches to LQG. We show how using real Ashtekar-Barbero variables might result in a different form of the MDR than when using the self dual Ashtekar variables. This is a remarkable result since the form of the MDR seems to depend on the choice of the Immirzi parameter. Although traditionally real variables have been more popular, some recent results have rekindled the usefulness of the self-dual variables \cite{SDSS,SDLQCP,jibr1,jibr3} and, thus, this seems an opportune moment to confront these two approaches with potentially observable predictions.

We study the case of the self dual variables in three different approaches. The primary strategy followed by us would be to look at spherically symmetric spacetimes and incorporate effects of holonomy corrections, based on the real and self dual varibles, in the hypersurface deformation algebra of such a system. Then we take its flat limit to derive the form of the MDR. Our holonomies are going to be always based on extrinsic curvature components, rather than the real-valued Ashtekar-Barbero connections or the self dual Ashtekar ones. In the approaches, where a signature-changing deformation function is allowed, we strictly restrict ourselves to the Lorentzian part of the spacetime and thus, for our purposes, do not consider more conceptual questions regarding the nature of the singularity-resolution. We would like to point out that the deformation function allowed in our case depends on both the radial variable and time, and is not spatially constant as has been studied earlier for derivation of a MDR from loop quantum cosmology \cite{LQCPoincare}. Additionally, we revisit the question of the signature change for self dual variables, within the framework in which quantum corrections have been implemented in this paper.

%We show a path to derive the form of the dispersion relation from the effective regime of Loop Quantum Gravity (LQG). Such an effective theory is given by the classical constraint equations where, though, \textit{holonomy corrections} are taken into account. These semi-classical effects modify the form of the hyper-surface deformation algebra. Such a deformation leaves trace in the Minkowski limit where a corresponding deformed Poincaré algebra arises. This allows us to derive a modified dispersion relation (MDR). Remarkably, the form of the MDR depends on the choice of the Immirzi parameter.

The paper is organized as follows. In Section \ref{II} we concentrate on the most-studied case of real $SU(2)$ Ashtekar-Barbero variables. Following Bengtsson \citep{Beng} we introduce the phase space reduced to spherical symmetry. After having replaced connections with their holonomies that are expected to account for quantum effects, the algebra of effective constraints is computed. For simplicity we adopt this effective scheme here, but is has been proven that the same results can be obtained by computing the expectation value of quantum operator constraints over coarse-grained spin network states (see e.g. \citep{BojoSwiderski,QRLG}). Then we restrict to the flat (Minkowski) spacetime limit by selecting suitable lapse and shift function. This allows us to find out a corresponding deformation of the Poincaré algebra, which has been first derived in \cite{LQGkPoincare} for real values of the Immirzi parameter. Finally, we find the form of the MDR requiring its invariance under deformed symmetries.  The same analysis is repeated in Section III for complex Ashtekar's variables. In particular, we compute holonomy corrections in three different quantization scheme: $SL(2, \mathbb{C})$ holonomies (\ref{naive}), which have to be regarded as a \textit{complexification} of the real ones, the \textit{analytic continuation} technique (\ref{AC}, and \textit{generalized holonomies} (\ref{GenHol}). For our purposes the main difference consists in the form of the implementation of holonomy corrections, which are responsible for the modification of the dispersion relation. In Section \ref{III} we discuss signature change in self-dual LQG. Conclusions are given in Section \ref{concl}.
We work in natural units.

\section{Real variables}\label{II}

LQG is a canonical approach to quantizing gravity, typically based on the real-valued Ashtekar-Barbero connection
\begin{equation}
A^{i}_{a} = \Gamma^{i}_{a}+\gamma K^{i}_{a}
\end{equation}
which are conjugate to the densitized triad $E^{a}_{i} = \sqrt{q}e^{a}_{i}$, with
\begin{equation}
\{ A^{i}_{a} (x) , E^{b}_{j}(y) \} = 8\pi G \gamma \delta^{b}_{a}\delta^{i}_{j}\delta^{(3)}(x-y)\,.
\end{equation}
Both $A^{i}_{a}$ and $E^{a}_{i}$ are functions defined on the spatial three dimensional manifold $\Sigma$.

It is possible to reduce the above variables to the spherically symmetric case  as has been shown in \cite{BojoSwiderski}. We provide a brief sketch of the procedure as follow. If $L_{i}$ are the rotational Killing vectors, we can obtain connections and triads which are invariant under rotations by solving the equation
\begin{equation}
\label{rotinveq}
\mathcal{L}_{L_{j}}E^{a}_{i} = -[T_{j},E^{a}_{i}] = -\epsilon_{ijk}\lambda_{j}E^{a}_{k}\,,
\end{equation}
where $T_{j}$ are the generators of $O(3)$, while $\lambda_{j}$ are just constants.

The solution of Eq. \eqref{rotinveq} is given by the following connections:
\begin{equation}
(A_{r}(r)\tau_{3}, A_{1}(r)\tau_{1}+A_{2}(r)\tau_{2}, A_{1}(r)\tau_{2}-A_{2}\tau_{1})\,,
\end{equation}
where $A_{r}, A_{1}, A_{2}$ are real functions which are canonically conjugate to $E^{r}, E^{1}, E^{2}$, while $\tau_{i} = -\frac{i}{2}\sigma_{i}$ are the $SU(2)$ generators, $r$ being the radial variable. The three manifold has been decomposed as $\Sigma = B\times S^{2}$, where $S^{2}$ are two-spheres of radius $r$ and $B = \mathbb{R}$.
Defining the angular connections and triads as
\begin{equation}
A_{\phi} := \sqrt{A_{1}^{2}+A_{2}^{2}}\,, \qquad E^{\phi} := \sqrt{E_{1}^{2}+E_{2}^{2}}\,,
\end{equation}
where from now on we  suppress the dependence on $r$. We also introduce  `internal directions' (on the $SU(2)$ tangent space)
\begin{eqnarray}
\tau^{A}_{\phi} &:=& \frac{A_{1}\tau_{2}- A_{2}\tau_{1}}{A_{\phi}}\,,\\
\tau^{\phi}_{E} &:=& \frac{E^{1}\tau_{2}-E^{2}\tau_{1}}{E^{\phi}}\,,
\end{eqnarray}
which allow us to define the `internal angles' $\alpha$ and $\beta$ via the relations
\begin{eqnarray}
\tau^{A}_{\phi} &=:& \tau_{1}\cos(\beta)+\tau_{2}\sin(\beta)\,,\\
\tau^{\phi}_{E} &=:& \tau_{1}\cos(\beta+\alpha) +\tau_{2}\sin(\beta+\alpha)\,.
\end{eqnarray}
Note that $A_{\phi}$ is not canonically conjugate to $E^{\phi}$, which is instead the momentum of the combination $A_{\phi}\cos\alpha = \gamma K_{\phi}$ (and thus conjugate to the angular extrinsic curvature component), i.e.:
\begin{equation}
\{A_{\phi}\cos\alpha(r), E^{\phi}(r') \} = \gamma G \delta(r-r')\,.
\end{equation}
The angular component of the extrinsic curvature $K_{\phi}$ can be read off from the relation $A_{\phi}^{2} = \Gamma_{\phi}^{2}+\gamma^{2}K_{\phi}^{2}$, where $\Gamma_{\phi} = -E^{r\prime}/(2E^{\phi})$ \footnote{The prime $\prime$ stands for the derivative with respect to the radial coordinate, i.e. $E^{r\prime} = \partial_{r}E^{r}$.}.

Assuming that the Gauss constraint has been solved classically\footnote{This allows us to reduce the phase space to two pair of canonical variables $(K_{r},E^r)$ and $(K_{\phi}, E^\phi)$.}, we can write the (spatial) diffeomorphism and the scalar (Hamiltonian) constraint respectively as:
\begin{eqnarray}
D[N^{r}] &=& \frac{1}{2G}\int_{B} dr N^{r}(2E^{\phi}K^{\prime}_{\phi} -K_{r}E^{r\prime})\,,\label{classmom}\\
H[N] &=& -\frac{1}{2G}\int_{B} dr N \left[ K_{\phi}^{2}E^{\phi}+2K_{r}K_{\phi}E^{r}\right. \nonumber\\
 & &\qquad \left.+(1-\Gamma_{\phi}^{2})E^{\phi} +2\Gamma_{\phi}^{\prime}E^{r}\right]\,,\label{classsc}
\end{eqnarray}
where we have used the definition $A_{r} = \Gamma_r +\gamma K_{r}$. At this point, the symplectic structure of the theory is given by the two Poisson brackets
\begin{eqnarray}
\{ K_{r}(r), E^{r}(r') \} &=& 2G\,\delta(r-r')\label{sym1}\\
\{ K_{\phi}(r), E^{\phi}(r') \} &=& G\, \delta(r-r')\label{sym2}
\end{eqnarray}
Given the above Eqs. \eqref{sym1}-\eqref{sym2} it is easy to compute the classical hypersurface deformation algebra as
\begin{eqnarray}
\{ D[N^{r}], D[N^{r'}] \} &=& D[N^{r}\partial_{r}N^{r'} -N^{r'}\partial_{r}N^{r}]\label{classhyp1}\\
\{ D[N^{r}], H[N] \} &=& H[N^{r}\partial_{r}N]\label{classhyp2}\\
\{ H[N], H[N^{'}] \} &=& D[g^{rr}(N\partial_{r}N^{'} -N^{'}\partial_{r}N)]\label{classhyp3}
\end{eqnarray}
where the  inverse of the spatial metric $g^{rr} = E^{r}/(E^{\phi})^{2}$.

In fact, in order to obtain Eqs. \eqref{classhyp1}-\eqref{classhyp2}-\eqref{classhyp3}, it is sufficient to use Eqs. \eqref{sym1}-\eqref{sym2} taking into account that the only non-vanishing Poisson brackets are those between a component of the extrinsic curvature and a derivative of the conjugate densitized triad (or vice-versa). However, by a way of example, we compute explicitly Eq. \eqref{classhyp1}  (a full derivation can be found, for instance, in \cite{CovSpherSymm, WDWSS})
\begin{widetext}
\begin{eqnarray}
\{ D[N^{r}], D[N^{r'}] \} &=& \frac{1}{4G^{2}}\int_{B} dr dr' N^{r}(r)N^{r'}(r')  \left[ \{ 2E^{\phi}(r)\partial_{r}K_{\phi}(r), 2E^{\phi}(r')\partial_{r'}K_{\phi}(r')  \} - \{ K_{r}(r)\partial_{r}E^{r}(r), K_{r}(r')\partial_{r'}E^{r}(r') \}\right] \nonumber \\
&=&  \frac{1}{4G^{2}}\int_{B} drdr' N^{r}(r)N^{r'}(r')\left[ 4GE^{\phi}(r) \partial_{r'}K^{\phi}(r')\partial_{r}\delta(r-r') -4GE^{\phi}(r') \partial_{r}K^{\phi}(r)\partial_{r'}\delta(r'-r)\right. \nonumber\\
& &  \hspace{2cm} \left. +2K_{r}(r)\partial_{r'}E^{r}(r') \partial_{r}\delta(r-r') - K_{r}(r')\partial_{r}E^{r}(r) \partial_{r'}\delta(r'-r)\right] \nonumber\\
&=& \frac{1}{2G}\int_{B} dr dr' \left[ 4(-\partial_{r}N^{r}(r)N^{r'}(r') E^{\phi}(r)\partial_{r'}K^{\phi}(r') + N^{r}(r)\partial_{r'}N^{r'}(r')E^{\phi}(r')\partial_{r}K^{\phi}(r))\right.\nonumber\\
& &\hspace{2cm} \left. +2(-\partial_{r}N^{r}(r)N^{r'}(r')K_{r}(r)\partial_{r'}E^{r}(r')  +N^{r}(r) \partial_{r'}N^{r'}(r')K_{r}(r') \partial_{r}E^{r}(r))\right]\delta(r-r')\nonumber\\
&=& \frac{1}{2G}\int_{B}dr (N^{r}(r)\partial_{r}N^{r'}(r) -N^{r'}(r)\partial_{r}N^{r}(r)) (2E^{\phi}(r)\partial_{r}K_{\phi}(r) -K_{r}(r) \partial_{r}E^{r}(r))\nonumber\\   &=& D[N^{r}\partial_{r}N^{r'}-N^{r'}\partial_{r}N^{r}]
\end{eqnarray}
\end{widetext}
The calculation of Eqs. \eqref{classhyp2}-\eqref{classhyp3} can be performed following the same steps.

Having set up our basics, we now want to study how (loop) quantum corrections deform the hypersurface deformation algebra. To this end we turn to the effective LQG theory by polymerizing the angular extrinsic curvature component:
\begin{equation}
\label{Rholocorr}
K_{\phi} \rightarrow \frac{\sin(K_{\phi}\delta)}{\delta}\,,
\end{equation}
where $\delta$ is related to some scale, usually $l_{Pl}$, as suggested, for instance, by the discrete spectrum of the area operator ($\delta$ is proportional to the square root of the minimum eigenvalue, or the `area gap' from LQG). Clearly, the classical regime is recovered in the limit $\delta \longrightarrow 0$\footnote{ The fact that zero does not belong to the spectrum of the area operator in LQG is precisely the input from the full theory which gives a nontrivial quantum geometrical effect.}. The above substitution \eqref{Rholocorr} can be justified as follows. In the quantum theory there is no well-defined operator corresponding to the Ashtekar-Barbero  connection $A^{i}_{a}$ on the LQG kinematical Hilbert space. Instead, in the loop representation, a well-defined object is the holonomy operator which are defined as parallel transport of the connection
\begin{equation}
\label{holo}
h_{\alpha}(A) = \mathcal{P}\exp(\int_{\alpha}\dot{e}^{a}A^{i}_{a}\tau_{i})\,,
\end{equation}
where $\mathcal{P}$ is the path-ordering operator and $\dot{e}^{a}$ is the three vector tangent to the curve $\alpha$. For our analysis are of particular interest the holonomies of connections along homogeneous directions, which simplify as
\begin{equation}
\label{homoholo}
h_{j}(A) = \exp(\mu A \tau_{j}) = \cos(\mu A) \mathbb{I}+\sin(\mu A)\sigma_{j}
\end{equation}
and do not require a spatial integration since they transform as scalars. In fact, so far  one knows only how to implement (local) holonomy corrections for connections along homogeneous directions (for a negative result concerning implementation of nonlocal (extended) holonomy corrections in spherical symmetry see \cite{SSnonlocalhol}). In our case, this is given by $\gamma K_{\phi}$ ( $= A_{\phi}\cos \alpha$):
\begin{eqnarray}
\label{angholo}
h_{\phi} ( r,\mu) &=& \exp(\mu A_{\phi}\cos\alpha \Lambda^{A}_{\phi})\nonumber\\
 &=& \cos(\mu \gamma K_{\phi}) \mathbb{I}+\sin(\mu \gamma K_{\phi}) \Lambda
\end{eqnarray}
In order to see how the replacement \eqref{Rholocorr} is implied by Eq. \eqref{angholo} one must take into account that the scalar constraint \eqref{classsc} is quantized by  utilizing the Thiemann trick $\sqrt{E^{r}}  \propto \{ K_{\phi}, V \}$ (where $V$ is the volume), whose quantum version contains the commutator $h_{\phi}[h^{-1}_{\phi},\widehat{V}] = h_{\phi}h^{-1}_{\phi}\widehat{V}- \widehat{V}h^{-1}_{\phi}\widehat{V}h_{\phi}$. (This is equivalent to regularizing the curvature of the connection by holonomies, with the minimum area being the `area gap' from LQG.) Using Eq. \eqref{angholo} one can easily see that products of holonomies are given by cosine and sine functions of $K_{\phi}$. Finally, it turns out that the resulting quantum or `effective' (since we are going to ignore operator ordering issues, which are not crucial to our goals) scalar constraint could be obtained simply making the replacement of Eq. \eqref{Rholocorr}. This justifies the following form of the effective Hamiltonian constraint
\begin{eqnarray}
\label{quantsc}
H^{Q}[N] &=& -\frac{1}{2G}\int_{B} dr N \left[ \frac{\sin^{2}(K_{\phi}\delta)}{\delta^{2}}E^{\phi}\right.\\
& &\quad \left. + 2K_{r}\frac{\sin(K_{\phi}\delta)}{\delta}E^{r} +(1-\Gamma_{\phi}^{2})E^{\phi}+2\Gamma_{\phi}^{'}E^{r}\right]\,.\nonumber
\end{eqnarray}
While the effective diffeomorphism constraint \eqref{classmom} remains undeformed since spatial diffeomorphism invariance translates into vertex-position independence in LQG, which is implemented directly at the kinematical level by unitary operators generating finite transformations\footnote{In fact, there is no well-defined infinitesimal quantum diffeomorphism constraint in LQG for the basis spin network states. Some progress in constructing it has been achieved in \cite{Varadarajan}.}.

Once again, it is straightforward to show that only the Poisson bracket between two Hamiltonian constraints is deformed due to the introduction of point-wise (since they act at the vertices of spin-networks only) holonomy corrections resulting in:
\begin{equation}
\label{defhda}
\{ H^{Q}[N], H^{Q}[N^{'}] \} = D[\cos(2\delta K_{\phi})g^{rr}(N\partial_{r}N^{'}-N^{'}\partial_{r}N)]
\end{equation}
while the other two Poisson brackets \eqref{classhyp1}-\eqref{classhyp2} remain unmodified \cite{CovSpherSymm, WDWSS}.

Next we wish to take the Minkowski (flat) limit of this deformed hypersurface deformation algebra, which is given by Eqs. \eqref{defhda}-\eqref{classhyp1}-\eqref{classhyp2}, with the aim of deriving the corresponding deformation of the Poincaré algebra. The latter will be used to find the modification to the dispersion relation. In order to reduce to the Minkowski spacetime limit (see Refs. \cite{DeformedGR1,LQGkPoincare}) it is necessary to restrict to linear lapse and shift functions, that correspond to linear coordinate changes, i.e.:
\begin{equation}
\label{minklim}
 N^{k}(x) = \Delta x^{k}+R^{k}_{i}x^{i}\quad N(x) = \Delta t+v_{i}x^{i}
\end{equation}
and, at the same time, to flat spatial hypersurfaces i.e. $g_{ij} \equiv \delta_{ij}$.  With these restrictions general diffeomorphisms reduce to the subset of Poincaré transformations. Then, it is possible to read off the commutators between the Poincaré generators from the hypersurface deformation algebra. To this end, let us make explicitly the case of rotations. They are generated by the momentum constraint $D[N^{i}]$, since they produce tangential deformations of the hypersurfaces, with shift vector given by $N^i = R^{i}_{l}x^{l} = \epsilon^{ijl}\varphi_{j}x_{l}$ (where $\epsilon^{ijl}$ is the Levi-Civita symbol and $\varphi_{j}$ stands for the angle of a rotation around the $j$ axis). This can be easily understood as follows. Let us introduce a local Cartesian frame on $g_{ij}$ and consider a rotation around the $z$ axis (i.e. we are choosing $j = 3$). Then, the rotated coordinates are obtained just adding $N^i = \epsilon^{i3l}\varphi_{3}x_{l}$ to the starting coordinates $(x,y,z)$. In fact, we have that $x'^i = x^i +N^i$ since in this way we find $x' = x - \varphi_3 y$, $y' = y+ \varphi_3 x$, and $z' = z$, as we could expect. Having proven that $D[N^{i}]$ accounts for rotations, let us derive the Poisson bracket between two Lorentz generators of infinitesimal rotations (i.e. $\{ J_l, J_j \}$) from the hypersurface deformation algebra. In light of the above discussion, this can be done by inserting $N^l = \epsilon^{lik}\varphi_{i1}x_k$ and $M^j = \epsilon^{jmn}\varphi_{m2}x_n$ into
\begin{equation}
\label{hypmom}
\{ D[N^{l}], D[M^{j}] \} = D[\mathcal{L}_{N^i}M^j]
\end{equation}
and, doing so, we obtain
\begin{eqnarray}
\mathcal{L}_{N^i}M^j &=& N^{i}\partial_{i}M^{j}-M^{i}\partial_{i}N^{j}\nonumber\\
 &=& \epsilon^{ilk}\varphi_{l1}x_k \epsilon^{jmn}\varphi_{m2}\delta_{ni} -\epsilon^{imn}\varphi_{m2}x_n\epsilon^{jlk}\varphi_{l1}\delta_{ki}\nonumber\\
 &=& (\delta_{lj}\delta_{km}-\delta_{lm}\delta_{kj})\varphi_{l1} \varphi_{m2}x_k \nonumber\\
 & &\hspace{1cm} -(\delta_{mj}\delta_{nl}-\delta_{ml}\delta_{nj})\varphi_{l1}\varphi_{m2}x_n\nonumber\\
 &=& \varphi_{j1}\varphi_{k2}x_k -\varphi_{l1}\varphi_{j2}x_l \nonumber\\
 &=& -\epsilon^{jlk}\epsilon_{lts}\varphi_{t1}\varphi_{s2}x_k = -\epsilon^{jlk} \varphi_{l3}x_k
\end{eqnarray}
This means that the right-hand side of Eq. \eqref{hypmom} (i.e. the result of combining two rotations) is still a momentum constraint that implements infinitesimal rotations by an amount $\varphi_{l3}x_k = \epsilon_{lts}\varphi_{t1}\varphi_{s2}x_k $ or, in other words, we have shown that $\{ J_l, J_j \} = \epsilon_{ljk}J_k$. Following the same line of reasoning, one can easily realize that $N^k = \Delta x^k$  corresponds to spatial translations, $N = \Delta t$ is a time translation by an amount $\Delta t$, and finally $N =    v_{i}x^{i}$ represents a boost along the $i$-axis. Then, plugging proper combinations of these lapse and shift into the hypersurface deformation algebra it is possible to regain the full Poincaré algebra just as we did for $\{ J_l, J_j \}$. Thus, we have shown that, in the classical theory, one recovers the standard Poincaré algebra by taking the flat (linear) limit of the algebra of constraints.

In presence of holonomy corrections from Eq. \eqref{defhda}, we expect to find a similar deformed version of the Poincar\'e algebra. However, our main difficulty lies in the fact that deformations in the hypersurface deformation algebra arises in the form of the structure function getting modified by a function of the phase space variables, while deformations at the level of the Poincar\'e algebra implies modification of the algebra generators. What we find for the case of holonomy modifications, specifically in the case of spherically symmetric models, is that these two can be related using the relation \cite{DeformedGR1}
\begin{equation}\label{byexcur}
\lambda P_{r} = \frac{1}{G}\frac{K_{\phi}}{\sqrt{|E^{r}|}} = 2\delta K_{\phi}
\end{equation}
with the choice $\delta = \frac{\lambda}{2G\sqrt{|E^{r}|}}$ ($\lambda$ being a constant usually set equal to the Planck length),  $\delta$ being the parameter appearing in the correction function Eq. \eqref{Rholocorr} (this choice of delta is rather well motivated from the point of view of holonomy corrections in LQG, from which $\delta$ should depend on the inverse square root from considerations of lattice refinement \cite{latticerefinement}). The above relation can be proved as follows. We start form the Brown-York momentum, which plays the role of the generator of local translations, \cite{BY}
\begin{equation}
\label{by}
P_a=-2\int_{\partial\Sigma} d^{2}x\frac{\sigma_{ab}}{\sqrt{h}}n_{l}\pi^{bl}
\end{equation}
where $\pi^{bl}$ is the gravitational momenta, $n_{l}$ the vector normal to the spatial slices $\Sigma = r = const$, and $\sigma_{ab}$ the metric of the two-boundary embedded in $\Sigma$. In particular, we are interested in computing the radial momentum $P_r$ so that the above expression reads
\begin{equation}\label{radialby}
P_r = -2\int  \frac{g_{rr}\sin\theta}{\sqrt{h}}\pi^{rr} d\theta d\phi = -\frac{8\pi g_{rr}\pi^{rr}}{\sqrt{h}}
\end{equation}
Recalling the form of the line element \cite{DeformedGR1}
\begin{equation}
ds^2 = \frac{(E^\phi)^2}{|E^r|}dr^2 + |E^r|(d\theta^2 + \sin^2\theta d\phi^2)
\end{equation}
it is immediate to compute the square root of the determinant of the three metric i.e. $\sqrt{h} = E^\phi \sqrt{|E^r|}\sin\theta$, that appears in Eq. \eqref{radialby}. Again, using the form of the line element and also the extrinsic curvature components
\begin{eqnarray}
K_{rr} &=& K_r \frac{E^\phi}{\sqrt{|E^{r}|}} ,\quad K_{\theta\theta} = K_\phi\sqrt{|E^{r}|} ,\nonumber\\
K_{\phi\phi} &=& K_\phi \sqrt{|E^{r}|} \sin^2\theta\,,
\end{eqnarray}
we calculate $\pi^{rr}$:
\begin{equation}
\pi^{rr} = \frac{\sqrt{h}}{16\pi G}(K^{rr}-Kh^{rr}) = -\frac{1}{8\pi G} K_\phi \frac{|E^r|}{E^\phi}\sin\theta\,,
\end{equation}
having used the fact that
\begin{equation}
K = g_{ij}K^{ij} = K_r\frac{\sqrt{|E^{r}|}}{E^\phi} + \frac{2K_\phi}{\sqrt{|E^{r}|}}\,.
\end{equation}
Finally, inserting all this quantities into Eq. \eqref{radialby}, we obtain
\begin{equation}
P_r = \frac{1}{G}\frac{K_{\phi}}{\sqrt{|E^{r}|}}
\end{equation}
that gives us exactly Eq. \eqref{byexcur}.

Given that the (angular) momentum generator is related to the deformation resulting from holonomy correction, it is straightforward to find the following deformation of the commutator between the radial boost $B_{r}$ and the energy $P_{0}$:
\begin{equation}
\label{defpa}
[B_{r}, P_{0}] = iP_{r}\cos(\lambda P_{r})
\end{equation}
where we have used the relation Eq. \eqref{byexcur}.

Following \eqref{defpa}, we find the following MDR \cite{LQGkPoincare}
\begin{equation}
\label{Rmdr}
P^{2}_{0} = 2(\frac{\lambda P_{r}\sin\lambda P_{r} +\cos\lambda P_{r}-1}{\lambda^{2}}) \simeq P^{2}_{r}-\frac{\lambda^{2}}{4}P^{4}_{r}
\end{equation}
where we have taken into account that the other commutation relation (i.e. $ [B_{r}, P_{r}]$ and $[P_{r}, P_{0}]$) remain undeformed. Given Eq. \eqref{Rmdr}, one can also derive an energy-dependent velocity for massless particles:
\begin{equation}
v (E) = \frac{d E(p)}{dp} \simeq 1-\frac{3}{8}\lambda^{2}E^{2}
\end{equation}
where, writing the above formula, we have substituted the symmetry generators $P_0,P_r$ with the corresponding conserved charges $E,p$ in Eq. \eqref{Rmdr}.

Thus, we have shown that LQG holonomy corrections produce a deformation of the Poincaré algebra in the flat regime. Consequently, there is a modification of the energy-momentum dispersion relation. In the next section we find that a different MDR is produced when the analysis is carried out using self-dual Ashtekar's variables.

\section{Self dual variables}

Recently, self dual Ashtekar variables have gained some traction due to some results in black hole thermodynamics in the LQG framework \cite{jibr,jibr1,jibr3}. It turns out that one can derive the correct form of the Bekenstein-Hawking formula for both static and rotating black holes without any fine tuning of the Immirzi parameter, in this formalism. Moreover, the (space-time) transformation property of the self-dual variables have an advantage over their real-valued counterparts in terms of description as a gauge field \cite{Samuel}.  Other interesting features of self dual variables, with respect to the algebra of hypersurface deformations applied to midisuperspace quantizations in LQG, have recently been discovered \cite{SDSS,SDLQCP}. Thus it is only natural to inquire if there are phenomenological consequences of looking into the MDR due to such variables which can distinguish them from the real Ashtekar-Barbero connection.

The aim of this Section is to derive the MDR using the same procedure adopted in Section \ref{II}, but now working with self dual connections. We analyse three different quantization scheme based on well-known procedures in the LQG literature \cite{SDLQCAC, wils,wiela,3DQG}. Section III is divided in three parts. The first contains the formulation of effective constraints with holonomy corrections of self dual connections obtained by complexifying real variables. With this first choice, the holonomies are evaluated in the fundamental representations of $SL(2,\mathbb{C})$ group  just as the real case of Section \ref{II} was based on the fundamental representation of $SU(2)$.  The second part makes use of self dual connections (i.e. $\gamma = \pm i$) by exploiting the recently introduced procedure of analytic continuation that uses the continuous  representations of $SU(1,1)$ as the symmetry group \cite{SDLQCAC}. The third treats the same issue but using the the tool of generalized holonomies, as used in \cite{wils, wils2}. We compute the deformed dispersion relation for each of these three possibilities. Our approach is to extract a holonomy correction function from each of these approaches, which we then use to polymerize the effective Hamiltonian constraint. In each case, we specialize to the spherically symmetric gravitational system so as to be able to use the framework developed in the previous Section \ref{II}. We emphasize that, for the derivation of the polymerization functions, the original work was done in a homogeneous LQC scenario. Our intent to transfer the correction function to the spherically symmetric case is to examine its effect on the deformation function, which is impossible to do in a strictly minisuperspace setting (since the spatial diffeomorphism constraint in that case is trivially zero). Also, we are not deriving new rigorous regularization schemes for these approaches applied to midisuperspace models, but rather mimicking the work done for the real-valued variables to make first contact with observations. We return to this point in the next Section \ref{III}. Like before, we shall only be concerned with the holonomy components along homogeneous directions, i.e. only point-wise holonomy corrections.

In the first two approaches, we obtain MDRs which are different from each other and also with respect to Eq. \eqref{Rmdr}. On the other hand, the last MDR (i.e. the one we find adopting the generalized-holonomy approach) coincide with Eq. \eqref{Rmdr}. This will lead us to claim that different quantization techniques used in LQG, although not necessarily having physically inequivalent flat limits, are sometimes distinguishable relying solely on phenomenological grounds.

\subsection{Fundamental $SL(2,\mathbb{C})$ holonomies}\label{naive}
Self-dual connections are given by
\begin{equation}
A^{i}_{a} = \Gamma^{i}_{a}\pm i K^{i}_{a}
\end{equation}
where the Immirzi parameter is,  thus, purely imaginary i.e. $\gamma = \pm i$. The main difference with respect to real-valued connections is that now the variables $A^{i}_{a}$ are no more in the adjoint representation of the $SU(2)$ group but they are elements of the non-compact group $SL(2,\mathbb{C})$. Following Thiemann \citep{Compl1,Compl2}, we can obtain the latter gauge group through a complexification of the former. This means that any element $A \in SL(2,\mathbb{C})$ can be written as \cite{wiel}
\begin{equation}
\label{complexif}
A = A^{i}\tau_{i}
\end{equation}
with $A^{i} \in \mathbb{C}$ and $\tau_{i}$ are the $SU(2)$ generators already introduced at the beginning of the previous section.

As a first-pass at the problem, we choose to work in the fundamental representation of $SL(2,\mathbb{C})$. This is not well-justified from the point of view of LQG since the functions obtained in this case would then naturally be unbounded. As a result, singularity-resolution is not possible for such a naive choice of the representation for the effective constraints. Nevertheless, theoretical premonitions aside, one is still allowed to do this without violating any of the gravitational restrictions. Thus we want to emphasize this case only to be a toy model; a sort of warm-up exercise in deriving MDRs for self dual variables.

For the purposes of our analysis, the crucial thing is that, in light of  Eq. \eqref{complexif}, the holonomy of the  angular complex connection $A_{\phi}\cos\alpha = \gamma i K_{\phi}$ is given by
\begin{equation}
\label{Cholo}
h_{\phi} ( r,\mu) = \exp(\mu \gamma K_{\phi} \Lambda^{A}_{\phi}) = \cosh(\mu K_{\phi}) \mathbb{I}-2\sinh(\mu  K_{\phi})\overline{\Lambda}\,,
\end{equation}
with $K_{\phi} \in \mathbb{R}$. As a consequence, following the same line of reasoning  from Section \ref{II}, we can introduce the following holonomy corrections
\begin{equation}
\label{Cholocorr}
K_{\phi} \rightarrow \frac{\sinh(K_{\phi}\overline{\delta})}{\overline{\delta}},.
\end{equation}
Thus, we find the following form for the effective Hamiltonian
\begin{eqnarray}
H^{Q}[N] &=& -\frac{1}{2G}\int_{B} dr N \left[ - \frac{\sinh^{2}(K_{\phi}\overline{\delta})}{\overline{\delta}^{2}} E^{\phi}\right.\\ & & \quad\left. + 2K_{r} \frac{\sinh(K_{\phi}\overline{\delta})}{\overline{\delta}}E^{r} +(\Gamma_{\phi}^{2}-1)E^{\phi}-2\Gamma_{\phi}^{'}E^{r}\right]\nonumber
\end{eqnarray}
where we have considered only the Euclidean part since the Lorentzian one disappears when working with a  purely imaginary Immirzi parameter (the reason being that the coefficient of the Lorentzian part is given by $(1+\gamma^2)$).

It is then straightforward to calculate the Poisson brackets between the quantum-corrected effective constraints, on evaluating which  one finds the following deformation to the hypersurface deformation algebra
\begin{equation}
\label{Cdefhda}
\{ H^{Q}[N], H^{Q}[N^{'}] \} = D[\cosh(2\overline{\delta} K_{\phi})g^{rr}(N\partial_{r}N^{'}-N^{'}\partial_{r}N)]\,.
\end{equation}
Clearly, it is different in form from the real-valued case due to the difference in holonomy correction functions. Once again, we take the Minkowski limit, following the steps outlined in Section \ref{II}. From Eq. \eqref{Cdefhda}, it follows
\begin{equation}
[B_{r}, P_{0}] = iP_{r}\cosh(\lambda P_{r})\,,
\end{equation}
and, finally, the corresponding MDR takes the form:
\begin{equation}
\label{Cmdr}
P^{2}_{0} = 2(\frac{\lambda P_{r}\sinh\lambda P_{r} -\cosh\lambda P_{r}+1}{\lambda^{2}}) \simeq P^{2}_{r}+\frac{\lambda^{2}}{4}P^{4}_{r}\,.
\end{equation}
This implies the energy-dependent velocity of particles on such a deformed Poincar\'e spacetime takes the form
\begin{equation}\label{Cvel}
v (E) = \frac{d H}{dp} \simeq 1+\frac{3}{8}\lambda^{2}E^{2}\,.
\end{equation}

As is evident from the form of the MDR, such an approach to self dual variables would clearly be distinguishable from the real Ashtekar-Barbero variables in its effect on the resulting violation of Lorentz symmetry. As mentioned above, one might argue that fundamentally this theory is radically different from the real-valued one in that the self dual approach considered here \textit{cannot} resolve the Schwarzschild singularity (or, equivalently the Big Bang singularity when applied to early universe cosmology) unlike the previous one. Although this is certainly correct, such purely theoretical considerations is impossible to directly verify since such (Planck scale) energy scales are way out of reach of conceivable observations. Even when looked at as a toy model, we provide a concrete path towards differentiating this approach from the real-valued one on phenomenological grounds before moving on to more realistic approaches towards implementing holonomy corrections using self dual variables, as described in the next part.

\subsection{Analytic continuation: $SU(1,1)$ holonomies}\label{AC}
Now we want to address once again the system of self dual spherically symmetric LQG by using a recently proposed procedure, namely an \textit{analytic continuation} from the real Immirzi parameter to the imaginary one\cite{jibr, SDLQCAC}. This recent proposal, originally proposed for LQC and black hole entropy calculations, puts the self dual variables on a much more rigorous footing. The approach is based on the principle that imaginary Immirzi parameter has to be used in combination with an analytic continuation of the spin $j$ representations to $j = -\frac{1}{2} + \frac{i}{2} s$ with $s \in \mathbb{R}$. The need for such a procedure can be briefly justified as follows (see e.g. Refs. \citep{jibr1,jibr2} for further details). Consider the eigenvalues of the area operator in LQG:
\begin{equation}
a_l = 8\pi l_P^2 \gamma \sqrt{j_l(j_l+1)}\,.
\end{equation}
If the Immirzi parameter is purely imaginary $\gamma = \pm i$, then, as one can realize by looking at the above expression, the area eigenvalues necessarily become imaginary. This would prevent the area operator from being a candidate observable even at the level of the kinematical Hilbert space. A heuristic manner to avoid this drawback is given by the following analytic continuation
\begin{equation}
j_l \rightarrow \frac{1}{2}(-1 + is)
\end{equation}
since it is immediate to realize that it implies
\begin{equation}
a_l \rightarrow 4\pi l_P^2 \sqrt{s_l^2 +1}
\end{equation}
In this way the spectrum of the area operator becomes continuous but it remains real. In the language of group theory this corresponds to turning from $SU(2)$ to $SU(1,1)$ representations\footnote{ We note that the dimension of the representation also gets a similar analytic continuation in a systematic procedure in this formalism; however, it is unimportant for our purposes here.}.

The expression of the field strength in terms of holonomies of homogeneous connections has been derived in Ref. \cite{jibr} for an arbitrary representation $s$ of the non-compact $SU(1,1)$ symmetry group. For our purposes here, it is of interest the fact that the result of Ref. \cite{jibr} corresponds to the following effective holonomy correction
\begin{equation}\label{ACholocorr}
K_\phi \rightarrow \frac{\sinh(\delta K_\phi)}{\delta}\sqrt{\frac{-3}{s(s^2+1)\sinh(\theta_\phi)}\frac{\partial}{\partial \theta_\phi}(\frac{\sin(s\theta_\phi)}{\sinh(\theta_\phi)})}
\end{equation}
where we have introduced the class angle $\theta_\phi$ defined as
\begin{equation}
\sinh(\frac{\theta_\phi}{2}) = \sinh^2 (\frac{\delta K_\phi}{2})
\end{equation}
We refer to Ref. \cite{jibr} for formal details. Although the form of the function obtained here is not very tractable, it has been shown that one has a non-singular quantum cosmological solution on implementing it \cite{SDLQCAC}. As a side note, we remark that the effective solution of this system is only known so far in the cosmological context and a full quantum theory is still beyond reach.

Plugging these holonomy corrections \eqref{ACholocorr} into the Hamiltonian constraint, a tedious but straightforward computation reveals that the hypersurface-deformation algebra is modified as follows
\begin{widetext}
\begin{eqnarray}\label{ACdeformed}
\{ H^Q[N], H^{Q}[N^{'}] \} &=& \frac{-3}{s(s^2+1)} D\left[\cosh(2\delta K_\phi)\left(\frac{1}{\sinh(\theta_\phi)} \frac{\partial}{\partial \theta_\phi}\left(\frac{\sin(s\theta_\phi)}{\sinh(\theta_\phi)}\right)\right) \right.\nonumber \\
& &\quad \left. +\frac{\sinh(2\delta K_\phi)}{\delta} \frac{\partial \theta_\phi}{\partial K_\phi}\left(-\frac{\cosh(\theta_\phi)}{\sinh^2(\theta_\phi)} \frac{\partial}{\partial \theta_\phi}\left(\frac{\sin(s\theta_\phi)}{\sinh(\theta_\phi)}\right) +\frac{1}{\sinh(\theta_\phi)}\frac{\partial^2}{\partial \theta^2_\phi}\left(\frac{\sin(s\theta_\phi)}{\sinh(\theta_\phi)}\right)\right) \right.\nonumber\\
& & \quad \left. + \frac{\sinh^2(\delta K_\phi)}{\delta^2}\frac{\partial^2 \theta_\phi}{2\partial K^2_\phi}\left(-\frac{\cosh(\theta_\phi)}{\sinh^2(\theta_\phi)} \frac{\partial}{\partial \theta_\phi}\left(\frac{\sin(s\theta_\phi)}{\sinh(\theta_\phi)}\right) + \frac{1}{\sinh(\theta_\phi)} \frac{\partial^2}{\partial \theta^2_\phi} \left( \frac{\sin(s\theta_\phi)}{\sinh(\theta_\phi)}\right)\right) \right.\nonumber\\
& &\quad \left. + \frac{\sinh^2(\delta K_\phi)}{\delta^2}\left(\frac{\partial \theta_\phi}{\partial K_\phi}\right)^2 \left(\frac{1}{\sinh^3(\theta_\phi)} \frac{\partial}{\partial \theta_\phi} \left(\frac{\sin(s\theta_\phi)}{\sinh(\theta_\phi)}\right) -\frac{\cosh(\theta_\phi)}{\sinh^2(\theta_\phi)}\frac{\partial^2}{\partial \theta^2_\phi}\left(\frac{\sin(s\theta_\phi)}{\sinh(\theta_\phi)}\right) \right)\right.\nonumber\\
& &\quad \left. + \frac{1}{2\sinh(\theta_\phi)} \frac{\partial^3}{\partial \theta^3_\phi}\left(\frac{\sin(s\theta_\phi)}{\sinh(\theta_\phi)}\right) g^{rr}(N\partial_{r}N^{'} - N^{'}\partial_{r}N)\right]
\end{eqnarray}
\end{widetext}
At this point it is possible to follow the same steps  we worked out in Section \ref{II} in order to derive the corresponding deformation of the dispersion relation in the flat regime. Adopting the standard notation used in Refs. \cite{DeformedGR1,LQGkPoincare}, we call $\beta(K_\phi)$ the deformation function appearing in the Poisson bracket involving scalar constraints in Eq. \eqref{ACdeformed}. It allows us to rewrite the above equation in an implicit but more compact form as follows:
\begin{equation}\label{defhdaAC}
\{ H^Q[N], H^{Q}[N^{'}] \} = D[\beta(K_\phi)g^{rr}(N\partial_{r}N^{'}-N^{'}\partial_{r}N)]
\end{equation}
Then, in light of Eq. \eqref{byexcur} that still holds true, we deduce that, in the flat spacetime limit, the commutator between the radial boost $B_r$ and the generator of time translations $P_0$ is deformed $[ B_r, P_0 ] = i \beta(P_r) P_r$. Guided by the findings obtained in Section II working with real connections, we propose the following \textit{ansatz} for the MDR:
\begin{equation}
\label{ACmdr}
P^2_0 = f(P_r)
\end{equation}
where one can easily check $f(P_r)$ has to satisfy the relation
\begin{equation}\label{implicitMDR}
f(P_r) = 2\int \beta(P_r) P_r dP_r
\end{equation}
in order to ensure the invariance under the deformed relativistic transformations implied by \eqref{defhdaAC}. Although we do not calculate the explicit form of the MDR in this case due to the complicated nature of the deformation function, we can still numerically plot its behaviour, as shown below (see Fig. (\ref{figmdr})). This would illustrate crucial features of its behaviour even without deriving its analytical form. From a phenomenological point of view, what is of interest is the leading non-trivial correction to the dispersion relation. It can be found by making a series expansion of $\beta$ of Eqs. (\ref{ACdeformed})-(\ref{defhdaAC}) for small values of $\delta \approx 0$. In this way, making use of Eq. (\ref{implicitMDR}), we find for Eq. (\ref{ACmdr}):
\begin{equation}
P^2_0 \simeq P^{2}_{r}+\frac{\lambda^{2}}{4}P^{4}_{r}\,.
\end{equation}
and for the group velocity
\begin{equation}
v (E) = \frac{d H}{dp} \simeq 1+\frac{3}{8}\lambda^{2}E^{2}\,.
\end{equation}
Notice that these expressions coincide with Eqs. (\ref{Cmdr})-(\ref{Cvel}), which refer to the case with $SL(2,\mathbb{C})$ holonomies. However, it is not difficult to realize that such a convergence is present only at the leading order. Then, at the next order, the MDR in the analytic continuation scheme picks up a negative correction term while the MDR for  $SL(2,\mathbb{C})$ holonomies is positive-definite (see Eq. (\ref{Cmdr})). This can be immediately understood looking at Fig. (\ref{figmdr})).

\subsection{Generalized holonomies}\label{GenHol}
In a series of recent papers \cite{wils,wils2}, another novel way of dealing with self dual Ashtekar variables has been proposed. It is based on the introduction of new fundamental variables which are called \textit{generalized holonomies}. They are defined as
\begin{equation}\label{genholo}
h_{\alpha}(A) = \mathcal{P}\exp(\int_{\alpha}\dot{e}^{a}iA^{i}_{a}\tau_{i})\,,
\end{equation}
where the fundamental difference with respect to standard holonomies of Eq. \eqref{holo} consists in an additional factor of $i$ multiplying the complex connection $A^{i}_{a}$. The main motivation for introducing these objects comes from the fact that, as shown in Ref. \cite{wils}, standard holonomies cannot be defined in the kinematical Hilbert space of LQC. Generalized holonomies retain some important properties. However, one of the major drawbacks is that they transform in a simple manner under gauge transformations, thereby loosing one of the pivotal characteristics of holonomies.

\begin{figure}[h!]
\centering
\includegraphics[width=3in]{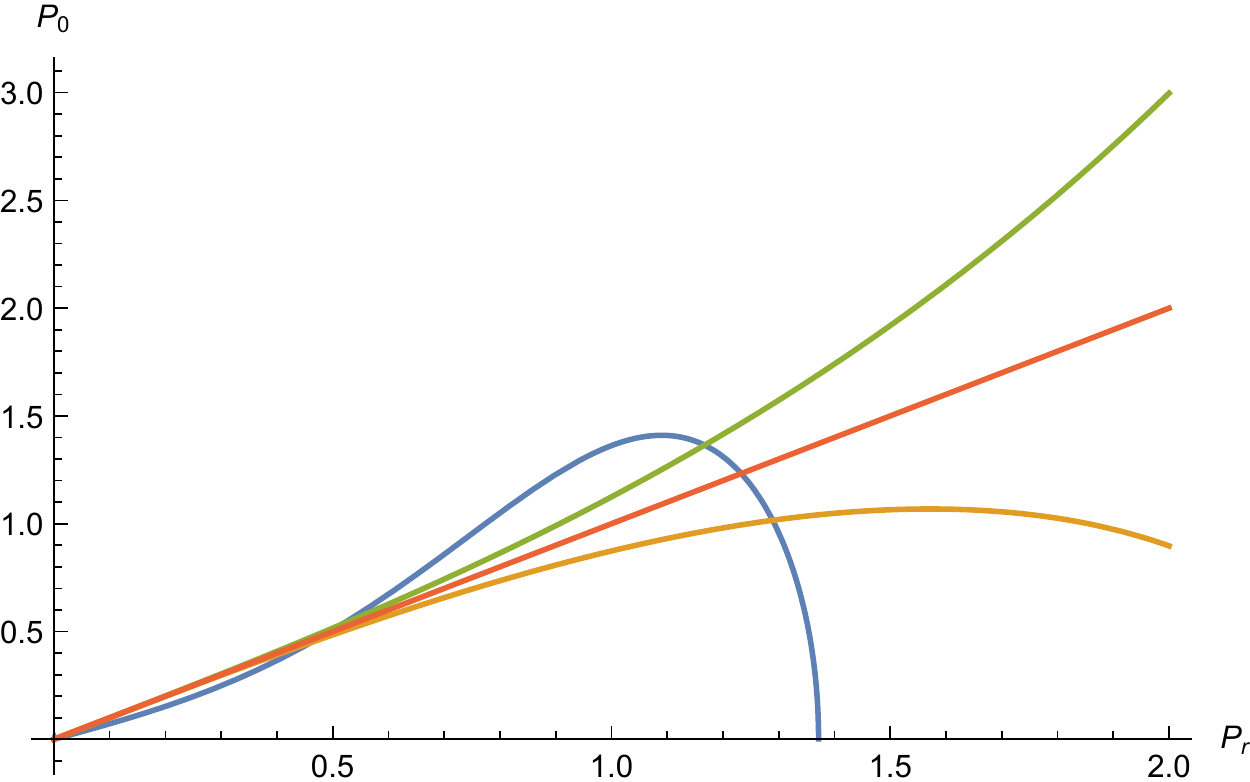}
\caption{
Behavior (for $0\leq P_{r}< 2$)
of the on-shell relations for massless particles ($m=0$) implied by four different mass Casimirs: the red line gives the usual special-relativistic dispersion relation, the orange line is the MDR obtained with both real \eqref{Rmdr} and generalized connections, the green line is the one given by Eq. \eqref{Cmdr}, and the blue line is the MDR in the analytic continuation case \eqref{ACmdr}. We set $\lambda \equiv 1$ and $s \rightarrow 0$.
}
\label{figmdr}
\end{figure}

Here we wish to seek which is the form of effective quantum corrections carried by generalized holonomies and, furthermore, how they affect the Poisson bracket $\{ H, H \}$. To this end let us consider a generic homogeneous complex connection, which we call $c(r)$, and his conjugated momentum $p(r')$ such that $\{ c(r), p(r') \} = i \delta(r-r')$. From Eq. \eqref{genholo} it follows that the holonomy of $c(x)$ is given by
\begin{equation}
h_{j}(c) = \exp(\mu c \tau_{j}) = \cosh(\mu c) \mathbb{I}+\sinh(\mu c)\sigma_{j}\,.
\end{equation}
If we take $c(r) = \gamma K_\phi(r) = i K_\phi(r)$ we can rewrite the above equation as
\begin{eqnarray}
h_{\phi} ( r,\mu) &=& \cosh(\mu i K_{\phi}) \mathbb{I}+\sinh(\mu i K_{\phi}) \sigma_\phi \nonumber\\
&=& \cos(\mu K_{\phi}) \mathbb{I}+\sin(\mu K_{\phi}) \Lambda\,,
\end{eqnarray}
which coincides exactly with Eq. \eqref{angholo}. This means that, for what regards holonomy corrections, the real case is equivalent to the self-dual case formulated in terms of generalized holonomies. In fact, in both cases, holonomy corrections yield the same substitution $K_\phi \rightarrow \sin(\delta K_\phi)/\delta$ in the effective Hamiltonian constraint (see Eq. \eqref{quantsc}). Finally, in light of the derivation of the Minkowski limit done in Section \ref{II}, we deduce that generalized-holonomy corrections produce the same deformation of the dispersion relation found in Eq. \eqref{Rmdr}. As a consequence, we claim that, relying on the form of the MDR, it is not possible to differentiate real effective LQG models from self-dual ones based on generalized holonomies. It would be necessary to figure out other observables that may allow to distinguish between holonomy corrections from real and self-dual variables, in the framework of \textit{generalized holonomies}.\\
In Fig. (\ref{figmdr}) we compare the MDRs found in self-dual variables using different quantization techniques with the MDR obtained for the real case, and also to the dispersion relation of special relativity.

\section{Signature-change for self dual variables?}\label{III}
The findings in this paper seem to be in apparent conflict with other recent results regarding self dual Ashtekar connections. Specifically, it has been shown that for spherically symmetric gravity \cite{SDSS}, the algebra of the holonomy-corrected constraints have the same form as the classical hypersurface-deformation algebra. This implies that the structure functions of the quantum-corrected constraint algebra does not pick up any modifications, for holonomies based on self dual connections. On the other hand, in this paper we show that self dual connections can also lead to deformations in the algebra when holonomy corrections are included in the algebra. The main source of difference stems from the fact that in \cite{SDSS} the holonomies are based on the Ashtekar variables, $A^i_a$, whereas in this paper they are implemented based on extrinsic curvature components. The Immirzi parameter is chosen to be $\pm i$ in both the approaches; however, it appears differently in the implementation of holonomy corrections. The main difference in the mathematical structure of the Hamiltonian constraint comes from the fact that the spin connection terms, which contains spatial derivative of the triad components\footnote{It has been shown that such terms are primarily responsible for deformations in the constraint algebra \cite{LQGworseBH}.}, do not arise when one works with the $\left(A , E\right)$ variables as the coefficients of these terms are proportional to $\left(1 + \gamma^2\right)$. However, when working with $\left(K , E\right)$ variables, such terms are re-introduced into the Hamiltonian when one expresses the self dual Ashtekar variable in terms of the extrinsic curvature component and the spin connection. Explicitly, the Hamiltonian constraint in our work has all the terms written in terms of $(K_r, K_\phi, E^r, E^\phi)$ and derivatives of the triad components. The dependence on the Immirzi parameter has been written explicitly wherever they appear (since it appears only in the form of $\gamma^2$, their effect is limited to a sign factor in front of some of the terms). On the other hand, the Hamiltonian constraint in \cite{SDSS} are formed out of the variables $(A_i, E^i),\; i=1,2,3$, and derivatives of the self dual connection components. This means the Immirzi parameter remains hidden implicit wherever components of the Ashtekar connection show up. For the approach taken in this article, the Immirzi parameter comes back through the implementation of the local correction functions. (An additional difference between the two approaches lies in the fact that the Gauss constraint is solved classically in our work, whereas it has been kept unsolved in \cite{SDSS} with an additional canonical pair of variables.) Thus the phase space of the two systems, although classically equivalent, are different in the two approaches. However, two systems which are classically equivalent \textit{can} give rise to quantum Hilbert spaces which are not unitarily-equivalent, this being a ripe example of it.

\begin{figure}[h!]
\centering
\includegraphics[width=3in]{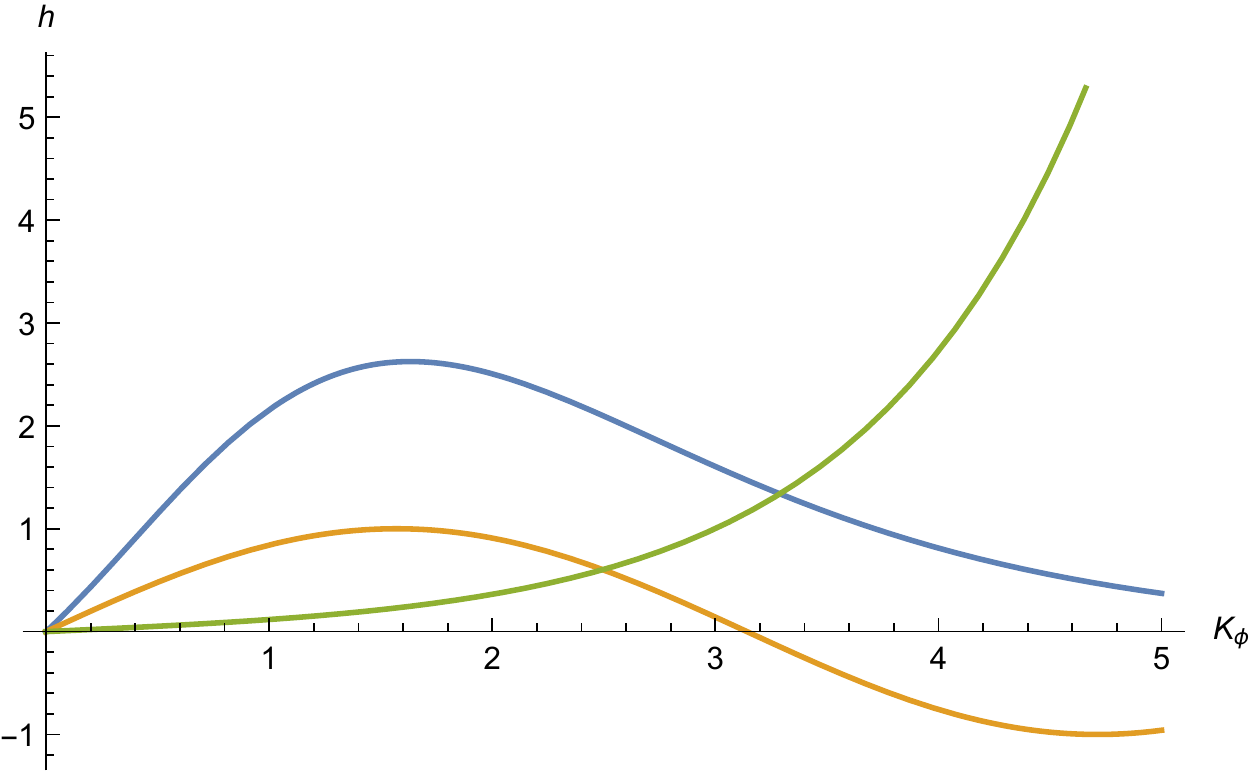}
\caption{
Behavior (for $0\leq\lambda K_{\phi}< 5$)
of holonomy-correction functions, i.e. $h(K_\phi)$, in the four different cases we analysed: the orange line stands for both real \eqref{Rholocorr} and generalized connections \eqref{genholo}, the green line is for complex connections given by Eq. \eqref{Cholocorr}, and the blue line represents the holonomy correction with analytic continuation \eqref{ACholocorr}. The discrete parameter $\delta$ is put equal to $1$. }
\label{figholo}
\end{figure}
\vspace{4cm}

\begin{figure}[h!]
\centering
\includegraphics[width=3in]{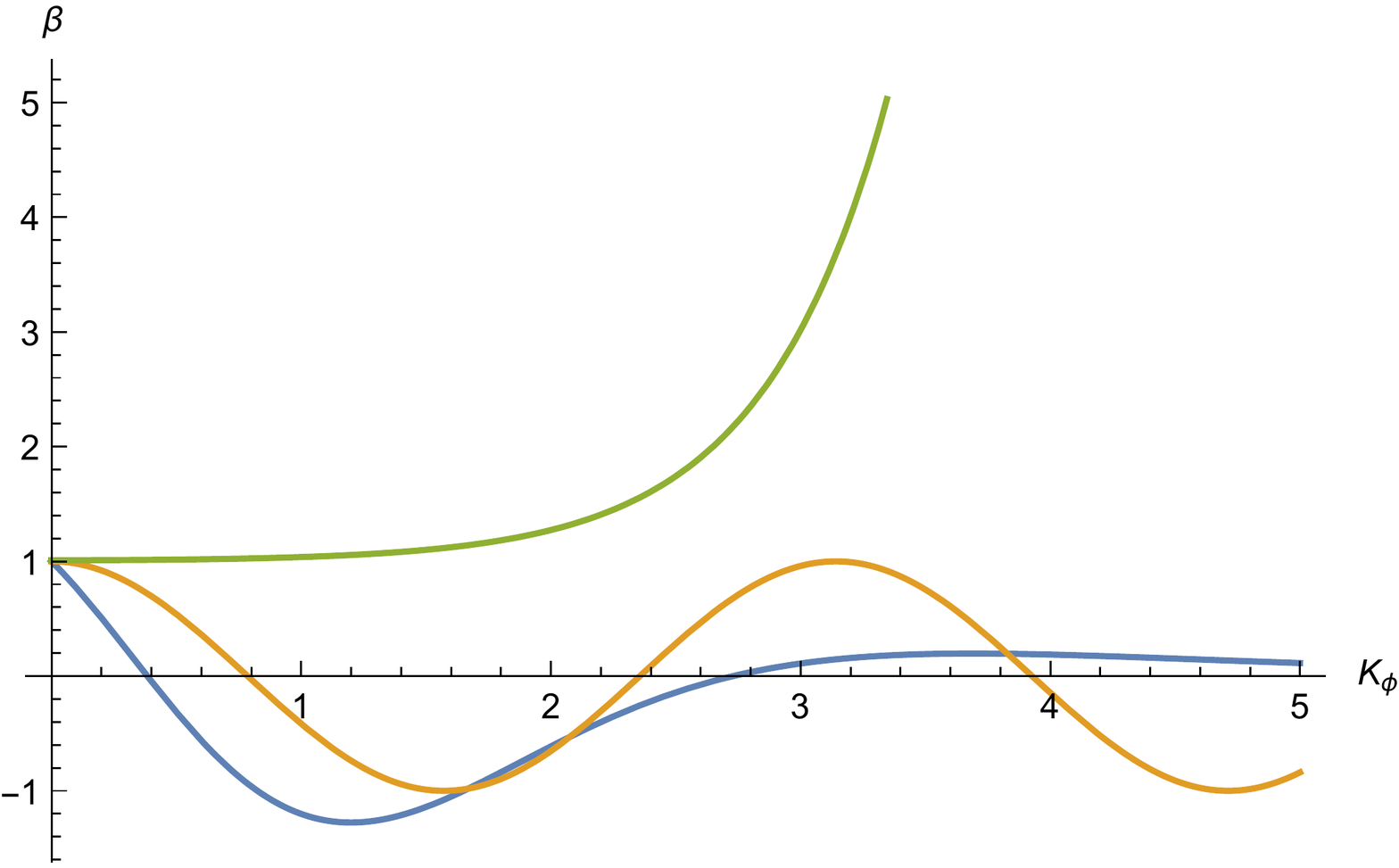}
\caption{
Behavior (for $0\leq\lambda K_{\phi}< 5$)
of the deformation function $\beta(K_\phi)$ in the four different cases we analysed: the orange line stands for both real \eqref{defhda} and generalized-connection cases, the green line is $\beta$ for complex $SL(2,\mathbb{C})$ connections \eqref{Cdefhda}, and the blue line is the one obtained using analytic continuation \eqref{defhdaAC}. The discrete parameter $\delta$ is put equal to $1$.
}
\label{figbeta}
\end{figure}

The next thing to investigate is the nature of the deformation function arising in the case of self dual variables, working in our approach. As is well-known in the real variables formalism, the deformation function appearing in the algebra changes sign in the deep quantum regime. This is known as signature change in the literature. The question now is whether the deformation function in the self dual case behaves in a similar manner or not. We examine this issue for each of the cases considered above individually. First, we explore the holonomy correction function for the naive ansatz of the holonomy calculated in the fundamental representation of the $SL(2,\mathbb{C})$ group, in Section \ref{naive}. In this case, the polymerization function as well as the deformation appearing in the algebra are both unbounded and do not exhibit any change of signature. This is easy to see from the analytic expressions for both these functions. Next, we look at the case for the generalized holonomies from Section \ref{GenHol}. In this case, both the holonomy correction function and consequently the deformation function are exactly the same as in the case of the real-valued variables, as has been described above. Thus we \textit{do} have signature changing deformations in this case. Finally, we examine the case of the self dual connection arising from an analytic continuation of the real Immirzi parameter, as in Section \ref{AC}. In this case, it has been shown that such a holonomy correction function results in singularity-resolution, when applied to a homogeneous and isotropic cosmological setting. We wish to emphasize the fact, already shown in \cite{SDLQCAC}, that this function has an upper bound which is of paramount importance for singularity-resolution. Since the classical singularity is resolved in the high curvature regime due to this upper bound, it follows that the deformation function, which is the second derivative of the holonomy correction function, turns \textit{necessarily negative} in those regimes. Thus the constraint algebra has the same sign as is required for Euclidean gravity and we have signature change for the analytic continuation example, provided one works with our implementation of incorporating quantum corrections. These analytical assertions are confirmed in the plot of both the holonomy correction functions as well as the deformation functions, as shown in Fig.(\ref{figholo}) and in Fig. (\ref{figbeta}).

\section{Conclusion}\label{concl}
Let us summarize our results as follows:
\begin{enumerate}
  \item We find that the choice of the Immirzi parameter, in particular, whether it is a real variable or a purely imaginary one can influence the form of the MDR due to a deformed Poincar\'e algebra. This, however, depends on the quantization scheme chosen for the self dual connection and only in the particular case of the `generalized holonomies', the MDR has the same form as for the real Ashtekar-Barbero connection.
  \item We further illustrate how self dual variables can also lead to the deformation of the hypersurface deformation algebra based on the implementation procedure of the holonomy corrections. However, it is worthwhile to point out that even if one takes the point of view that the self dual variables \textit{do not} deform the algebra, as in \cite{SDSS}, that does not change our central result that the MDR due to them are different from the case of the real-valued ones. In this case, one gets the familiar dispersion relation on Minkowski space for the self dual variables while a deformed one for the real ones.
  \item Additionally, we illustrate the nature of the deformation functions for the self dual variables, while proceeding with the different quantization schemes. While both the popular schemes chosen in LQG lead to a signature change, the more naive ansatz does not have a change in sign. However, this shows a more general trend that whichever quantization scheme leads to singularity-resolution within LQG, essentially also leads to a signature change in the deep quantum regime.
\end{enumerate}

We have taken the first steps towards confronting LQG with actual observable consequences, in how the local Poincar\'e symmetry might get deformed in such a quantum gravity theory. We wish to calculate other such observables which would be able to capture phenomenological signatures of LQG (see \cite{dimLQG} for a preliminary step taken by one of us toward this direction). We are also looking into the effects of other types of quantization schemes employed within LQG, depending on the representation of the internal gauge group, on the MDR of particles on a deformed (non-classical) Minkowski spacetime.
%
%In summary, we have shown that quantum holonomy corrections are sensitive to the choice of the Immirzi parameter, depending on the quantization scheme one choses. In particular, they get different functional forms if one chooses to work with real connections i.e. $\gamma \in \mathbb{R}$ or puts $\gamma = \pm i$. This reflects the fact that in the former case the Ashtekar's variables $A^{i}_{a}$ (and hence their holonomies) belong to the $SU(2)$ compact group, while the latter corresponds to $A^{i}_{a} \in SL(2, \mathbb{C})$. Interestingly, these quantum corrections affect the hypersurface deformation algebra in such a way that, in the Minkowski limit, a deformed Poincaré algebra comes out. Remarkably, the real and the self-dual cases produce different forms of the MDR and, thus, it could be possible to discriminate the real connections from the complex ones through the experimental tests of the MDR.

\bibliographystyle{apsrev4-1}
\bibliography{refs}

\end{document}